\documentclass{article}

\usepackage[final, nonatbib]{neurips_2024}

\usepackage[utf8]{inputenc} 
\usepackage[T1]{fontenc}    
\usepackage{hyperref}       

\usepackage{url}            
\usepackage{booktabs}       
\usepackage{amsfonts}       
\usepackage{nicefrac}       
\usepackage{microtype}      
\usepackage{xcolor}         
\usepackage{amsmath}
\usepackage{multirow} 
\usepackage{enumitem}

\usepackage[
style=ieee, 
uniquelist=false,
mincitenames=1,
maxcitenames=2,
minbibnames=1,
maxbibnames=2,
url=false,
doi=false,
eprint=false, 
backend=bibtex]{biblatex}

\usepackage{wrapfig}

\newcommand{\inlineSubsection}[2]{
  \refstepcounter{subsection} 
  \noindent\textbf{\thesubsection\ #1}\label{#2}
}

\newcommand{\ctx}{\textit{ctx} }
\newcommand{\clapa}{$\textit{CLAP}_{A} $}
\newcommand{\clapt}{$\textit{CLAP}_{T} $}
\newcommand{\uncond}{$\textit{uncond} $}

\title{Improving Musical Accompaniment Co-creation via Diffusion Transformers}

\author{
  Javier Nistal$^1$ \\
  \And
  Marco Pasini$^{2}$\thanks{This work is supported by the EPSRC UKRI Centre for Doctoral Training in Artificial Intelligence and Music (EP/S022694/1) and Sony Computer Science Laboratories Paris.} \\
  \\
  $^{1}$Sony Computer Science Laboratories, Paris \\
  $^{2}$Queen Mary University of London
  \And
  Stefan Lattner$^1$ \\
}

\begin{document}

\maketitle

\begin{abstract}
Building upon Diff-A-Riff, a latent diffusion model for musical instrument accompaniment generation, we present a series of improvements targeting quality, diversity, inference speed, and text-driven control. First, we upgrade the underlying autoencoder to a stereo-capable model with superior fidelity and replace the latent U-Net with a Diffusion Transformer. Additionally, we refine text prompting by training a cross-modality predictive network to translate text-derived CLAP embeddings to audio-derived CLAP embeddings. Finally, we improve inference speed by training the latent model using a consistency framework, achieving competitive quality with fewer denoising steps. Our model is evaluated against the original Diff-A-Riff variant using objective metrics in ablation experiments, demonstrating promising advancements in all targeted areas. 
Sound examples are available at: \url{https://sonycslparis.github.io/improved_dar/}.
\end{abstract}

\section{Introduction}

Deep generative models for audio are rapidly gaining traction due to their potential to transform music creation and audio manipulation. Recent advancements \cite{musiclm, musicgen, stable_audio, stable_audio_2, stable_audio_open, audioldm2} anticipate a future where humans and AI collaborate seamlessly to expand artistic expression. However, important challenges remain before these models can be fully integrated into professional music production workflows, including issues with audio quality, resource-intensive generation processes, and limited control mechanisms.

In response to these challenges, Diff-A-Riff \cite{diffariff} was recently introduced as a model specifically designed for music production purposes. Leveraging a Latent Diffusion Model \cite{rombach_high-resolution_2022} and a Consistency Autoencoder (CAE) \cite{music2latent}, Diff-A-Riff can generate high-quality, pseudo-stereo, single-instrument accompaniments that adapt to different musical contexts—for example, generating a guitar track conditioned on a mixture of drums and bass. It also provides control through text prompts and audio style references using CLAP embeddings \cite{clap}. While Diff-A-Riff represents a significant step toward AI-assisted music co-creation tools, opportunities for improvement remain, particularly in achieving true stereo output with enhanced quality, computational efficiency, and better text-driven generation capabilities.

In this work, we present a series of enhancements to Diff-A-Riff, focusing on three key areas: quality, speed, and control. First, we enhance audio quality by upgrading the autoencoder to a stereo-capable model with improved fidelity \cite{m2l2} and transitioning from a latent diffusion U-Net to a Diffusion Transformer (DiT) architecture \cite{dit}. As shown in our experiments, these improvements generally lead to more diverse and higher-quality audio generation. 
Second, we refine the model's text-to-audio capabilities by training a separate diffusion model to mitigate the modality gap inherent in CLAP space \cite{clap_gap_orig}, i.e., reducing the distance between audio-derived and text-derived CLAP embeddings. By incorporating this model as an interface between text prompts and Diff-A-Riff, we experimentally show improved generation quality and a better resemblance of the audio with the intended prompts.
Finally, to improve inference speed, we introduce consistency training to the latent model, maintaining competitive audio quality while enabling five-step generation. We rigorously evaluate our enhanced model against the original Diff-A-Riff using objective metrics in an ablation study. The results demonstrate improvements across the targeted areas, highlighting the potential of our improved model for practical music production applications.

\vspace{-2mm}
\section{Related Work}
\label{sec:related_work}
\vspace{-2mm}
Autoregressive models trained on waveform samples \cite{wavenet, samplernn} produce high-fidelity audio but are computationally expensive. GANs \cite{gan} and VAEs \cite{vae} offer faster inference but struggle with long-term dependencies and complex musical structures \cite{drumganvst, rave}. Denoising Diffusion Models have been applied to audio generation \cite{crash} but tend to be slow due to their iterative denoising process. 

Recently, models trained on compressed audio representations via autoencoders have shown promise for efficiently modeling long-form music. Autoregressive models using discretized VQ-VAE representations \cite{vqvae} allow for handling long-term structures and multi-modal inputs \cite{jukebox, musiclm, musicgen}. Latent Diffusion Models, which operate in continuous latent spaces, achieve competitive quality at high sample rates \cite{stable_audio, stable_audio_2, stable_audio_open}. Hybrid approaches \cite{seed_music} show promise for audio generation and editing.

Control mechanisms beyond text prompting are being explored to enhance precision in music production. Time-varying parameters \cite{jukedrummer, mcontrolnet} and latent manipulations in text-audio spaces \cite{musicmagnus, manor} offer finer control, while inference-time optimization \cite{ditto, ditto2} and guidance strategies \cite{cmp} improve model control. Conditioning on audio signals has been effective for style transfer (e.g., melody or timbre \cite{musicgen, bassnet2}) and accompaniment generation \cite{diffariff, singsong, stemgen, DrumNet, bassnet, bassnet2}. Recent efforts combine music generation and source separation to create individual stems \cite{postolache, multisourceddm, comp_rep, tornike}.


\vspace{-2mm}
\section{Background}\label{sec:background}
\vspace{-2mm}
\inlineSubsection{Music2Latent2 (M2L2)}{sec:m2l2}\footnote{Paper is under review. We provide an anonymized version at \href{https://drive.google.com/file/d/1ovp-M4YAvQhVemEYi5TWCtcXh8zvykYK/view?usp=sharing}{this link}.} introduces an audio autoencoder designed to achieve high compression rates while preserving audio fidelity. 
M2L2 employs an autoregressive consistency model trained with causal masking to handle arbitrary audio lengths by processing the input in consecutive chunks. The model consists of three components: an encoder that extracts so-called \emph{summary embeddings}--unordered latent representations capable of capturing global features of audio chunks-- from the input complex Short-Time Fourier Transform (STFT) spectrogram, a decoder that produces upsampled audio features, and a consistency model that reconstructs the original spectrogram based on the features from the decoder.
During inference, M2L2 uses a two-step decoding procedure, refining the generated audio by reintroducing noise into previously decoded segments. Compared to Music2Latent, which is used in Diff-A-Riff \cite{diffariff}, M2L2 supports stereo inputs and outputs while achieving superior reconstruction quality at double the compression ratio.

\inlineSubsection{Diffusion Transformers (DiT)}{sec:dit} replace the U-Net backbone used in image and audio-based diffusion models with a transformer-based architecture. DiT works by dividing the input representation into patches and processing them sequentially through transformer blocks. DiT uses Adaptive Layer Normalization (AdaLN), calculating normalization parameters from embeddings of diffusion timesteps and class labels. The AdaLN-Zero initialization strategy is shown to improve training stability and results further.
In this work, we train a Latent DiT \cite{sd3, snap_audio} on sequences of latent embeddings from Music2Latent2. Latent DiTs were recently shown to generate long-form music with high resolution \cite{stable_audio_open}.

\inlineSubsection{Consistency Models (CMs)}{sec:cm} \cite{consistency_models, improved_consistency_models} are a new type of generative model capable of producing high-quality samples in a single forward pass, bypassing the need for adversarial training or iterative sampling. They learn a mapping from noisy to clean data via a probability flow ODE \cite{ddim}. Given noise level $\sigma$, the consistency function $f$ transforms a noisy sample $x_\sigma \sim p_\sigma(x)$ into a clean sample $x \sim p_{data}(x)$ using $f(x_\sigma, \sigma) \mapsto x$, typically parameterized as a neural network $f_{\theta}(x_\sigma, \sigma)$.
To ensure $f_{\theta}(x, \sigma_{min}) = x$, where $\sigma_{min}$ is the minimum noise level, CMs are expressed as:
\[
f_{\theta}(x_\sigma, \sigma) = c_{skip}(\sigma)x_\sigma + c_{out}(\sigma)F_{\theta}(x_\sigma, \sigma),
\]
with $F_{\theta}$ as a neural network, and $c_{skip}(\sigma)$, $c_{out}(\sigma)$ as differentiable functions satisfying $c_{skip}(\sigma_{min}) = 1$ and $c_{out}(\sigma_{min}) = 0$. 
Training involves discretizing the ODE over noise levels $\sigma_1, \dots, \sigma_N$ and minimizing the consistency loss:
\[
\mathcal{L}_{CM} = \mathbb{E} [\lambda(\sigma_i, \sigma_{i+1})d(f_{\theta}(x_{\sigma_{i+1}}, \sigma_{i+1}), f_{\theta^-}(x_{\sigma_i}, \sigma_i))],
\]
where $d(x,y)$ is a distance metric, $\lambda(\sigma_i, \sigma_{i+1})$ scales the loss, and $f_{\theta^-}$ is a teacher network. After training, a sample $x$ is generated in one step from noise $z \sim \mathcal{N}(0, I)$ using $x = f_{\theta}(z, \sigma_{max})$.

\vspace{-2mm}
\section{Method}\label{sec:method}
\vspace{-2mm}
\inlineSubsection{Dataset}{sec:data}. We use a proprietary dataset of over 20,000 multi-track recordings across various music genres and instruments, with a sample rate of 48 kHz. During training, we select a target accompaniment track (excluding vocals) and mix a random subset of remaining tracks to create the music context (\ctx), following the same approach as in Diff-A-Riff \cite{diffariff}. We then divide training samples into 10-second windows. This process yields 1 million training pairs. A validation set is derived similarly from 2,000 recordings.

\inlineSubsection{Experiments}{sec:experiments} 
We conduct an ablation study to evaluate the impact of the modifications on Diff-A-Riff's original setting. The experiments focus on the following enhancements:

\begin{itemize}[left=0pt]
    \item \textbf{Autoencoder Improvement (\textit{M2L2})}: We replace the original Music2Latent autoencoder used in Diff-A-Riff with Music2Latent2 \cite{music2latent}, a state-of-the-art audio autoencoder featuring a transformer-based architecture and an autoregressive decoding scheme. Music2Latent2 supports stereo input and output, providing higher reconstruction quality at a 128× compression ratio—double that of the original Music2Latent—while maintaining the same latent dimensionality.
    \item \textbf{Architecture Transition (DiT)}: We shift from a convolutional U-Net to a Diffusion Transformer (DiT) \cite{dit} with depth-wise convolutions after Q, K, V projections in self-attention\cite{primer}. The DiT is initialized with a dimension of 1024, MLP multiplier of 4, 4 heads, and 18 layers. We use 512-dimensional sinusoidal embeddings \cite{transformer} for noise levels, fed into AdaLN layers \cite{dit}. The model has around 280 million parameters (150 million less than Diff-A-Riff).

    \item \textbf{Consistency Training for Faster Inference (\textit{C-DiT})}: To enhance inference speed, we introduce consistency training to the latent model, following the framework in \cite{improved_consistency_models}. We use continuous noise levels and an exponential schedule for the consistency step, as shown in \cite{music2latent}. The remaining parameters are unchanged from the ones used for EDM training. Consistency training allows the model to generate high-quality audio with a few inference steps, significantly reducing computational overhead. We use 5 inference steps, which already improves quality over Diff-A-Riff (see Sec.\ref{sec:results}).
    
    \item \textbf{Bridging the Modality Gap ($\textit{CLAP}_{\beta}$)}: To enhance the model's responsiveness to text prompts, we address the modality gap between audio-derived and text-derived CLAP embeddings \cite{clap, clap_gap, clap_gap2, clap_gap_orig}. This gap seems to arise primarily due to the contrastive objective \cite{contrastive_gap} and produces text and audio embeddings to lie in disjoint manifolds. To enable flexible sampling of multiple audio variations for a single text prompt—and to avoid the costly retraining of CLAP with new labels—we propose training a diffusion-based MLP model to reduce this gap. Leveraging a set of unstructured, human-annotated tags paired with our multi-track dataset (see Sec.~\ref{sec:data}), we train the MLP to predict audio-derived from text-derived CLAP embeddings, following a similar approach to \cite{dalle2}. The MLP is initialized with 1024 hidden units and uses eight residual dense blocks with Adaptive Layer Normalization (AdaLN) layers to integrate conditional text embeddings from CLAP. This setup effectively bridges the modality gap, improving generation quality for both text-prompted and unconditional audio CLAP embedding generation while allowing flexible sampling without requiring modifications to the CLAP model itself.

\end{itemize}

\inlineSubsection{Implementation Details}{sec:implementation}
We train all models for 1 million iterations using AdamW \cite{adamw}, with $lr=1e-4$, $\beta_1 = 0.9$, $\beta_2 = 0.999$. We use a linear warmup during the first 1,000 steps, followed by a cosine decay until $lr=0$ is reached. U-Net-based models are trained with a batch size of $256$, while DiT-based models use a batch size of $128$ for diffusion training and $16$ for consistency training. Other details follow the same methodology as in Diff-A-Riff. We address a key issue of information leakage in Diff-A-Riff, where the CLAP embedding is derived from the same audio segment as the target. Instead, we extract the CLAP embedding from a random segment of the same target.

\inlineSubsection{Evaluation}{sec:eval}
 We perform objective evaluations to assess the modifications made to Diff-A-Riff's original configuration \cite{diffariff}. Several metrics are used in this evaluation: \textit{Kernel Distance} (KD)~\cite{kid} and \textit{Fréchet Audio Distance} (FAD)~\cite{fad} assess audio quality while \textit{Density} and \textit{Coverage}~\cite{den_cov} measure fidelity and diversity. The Accompaniment Prompt Adherence (APA) \cite{apa} metric evaluates how well the generated accompaniment aligns with the given context. 
 
 The Clap Score (CS) \cite{Make-An-Audio} typically quantifies cross-modality similarities between individual pairs of text and audio in the CLAP space \cite{clap}. We refer to this metric as CS$_{AT}$. Additionally, we also report the \emph{intra}-modality (CS$_{AA}$). In both cases, CS is computed based on generated data projected back into the CLAP space and the respective ground truth \emph{embedding}. For \clapa\  conditioning, the ground truth for CS$_{AA}$ is the reference audio, and the ground truth for CS$_{AT}$ is the caption of the reference audio (which is not used for conditioning in the \clapa\ case). For \clapt\ conditioning, the ground truth for CS$_{AA}$ is the audio whose caption was used for \clapt\ conditioning, and the ground truth for CS$_{AT}$ is the actual caption used for computing the \clapt\ conditioning. For context-only conditioning (\ctx), the ground truth for CS$_{AA}$ is the audio stem that was originally part of the context but was removed and for CS$_{AT}$, it is the caption of that removed stem.

 All metrics are calculated by averaging five batches of 1000 candidate samples. We use CLAP \cite{clap} as the embedding space for metrics that compare distributions (like KD and FAD) using a reference set of 5,000 real audio examples.

\vspace{-2mm}
\section{Results}\label{sec:results}
\vspace{-2mm}
Results are summarized in Table~\ref{tab:results}, where we compare the performance of our enhanced models to the original Diff-A-Riff under various conditioning settings. 

\begin{table*}[t]

 \begin{center}
 \begin{footnotesize}
 \begin{tabular}{ccccccccc}
  \toprule
   & Inputs & $\downarrow$ KD$^a$ & $\downarrow$ FAD & $\uparrow$ \textit{Cov.}$^b$ & $\uparrow$ \textit{Den.}$^b$ & $\uparrow$ APA & $\uparrow$ CS$_{AA}$ & $\uparrow$ $CS_{TA}$\\

  \midrule
    {real} & Original acc. & 0.00 & 0.01 & 0.14 & 0.79 & 0.99 & - & 0.39 \\
    \midrule
    \textit{$\downarrow$ bound}       & - & 6.94$^c$ & 1.65$^c$ & 0.00$^c$ & 0.00$^c$ & 0.14$^d$ & - & 0.13$^d$ \\
    \midrule
    \midrule
    \multirow{6}{*}{\textit{Diff-A-Riff}}           
        &  \clapa, \ctx   & 1.48 & 0.42 & 0.07 & 0.42 & 0.85 & 0.53 & 0.13\\
        &  \clapt, \ctx   & 1.59 & 0.46 & 0.08 & 0.52 & 0.96 & 0.47 & \textbf{0.29}\\
        & \ctx            & 2.19 & 0.63 & 0.16 & 2.80 & 0.33 & 0.25 & 0.03\\
        & \clapa          & 1.55 & 0.44 & 0.07 & 0.46 &   -  & 0.51 & 0.17\\
        & \clapt          & 1.55 & 0.45 & 0.07 & 0.46 &   -  & 0.46 & \textbf{0.29}\\
        & \uncond         & 2.46 & 0.70 & 0.12 & 3.46 &   -  & 0.23 & 0.02\\
    \midrule
    \multirow{6}{*}{\textit{+M2L2}} 
        &  \clapa, \ctx   & 1.02 & 0.35 & 0.52 & 4.59 & 0.89 & 0.52 & 0.15\\
        &  \clapt, \ctx   & 1.10 & 0.36 & 0.48 & 5.89 & 0.96 & 0.47 & 0.19\\
        & \ctx            & 0.99 & 0.34 & 0.61 & 7.07 & 0.88 & 0.41 & 0.09\\
        & \clapa          & 1.11 & 0.38 & 0.44 & 4.13 &   -  & 0.49 & 0.13\\
        & \clapt          & 1.10 & 0.38 & 0.45 & 5.06 &   -  & 0.45 & 0.16\\
        & \uncond         & 1.03 & 0.36 & 0.50 & 6.00 &   -  & 0.39 & 0.08\\
    \midrule
    
    \multirow{6}{*}{\textit{+DiT}}                 
        &  \clapa, \ctx   & 0.90 & 0.33 & 0.73 & 5.94 & \textbf{1.00} & \textbf{0.53} & 0.18\\
        &  \clapt, \ctx   & 0.96 & 0.34 & 0.57 & 7.24 & \textbf{1.00} & 0.47 & 0.19\\
        & \ctx            & \textbf{0.88} & \textbf{0.31} & 0.62 & 7.53 & 0.99 & 0.42 & 0.11\\
        & \clapa          & 1.04 & 0.38 & 0.61 & 5.39 &   -  & 0.49 & 0.15\\
        & \clapt          & 1.04 & 0.38 & 0.57 & 7.29 &   -  & 0.44 & 0.15\\
        & \uncond         & 1.00 & 0.35 & 0.65 & 6.75 &   -  & 0.39 & 0.08\\

    \midrule
    \multirow{4}{*}{\textit{+$\textit{CLAP}_{\beta}$}}
        &  \clapa$^{*}$, \ctx   & 0.93 & 0.33 & 0.65 & 6.51 & 0.99 & 0.50 & 0.15\\
        &  \clapt, \ctx         & \textbf{0.88} & 0.32 & \textbf{0.92} & \textbf{11.1} & 0.87 & 0.49 & 0.19\\
        & \clapa$^{*}$          & 1.00 & 0.35 & 0.58 & 6.38 &   -  & 0.48 & 0.14\\
        & \clapt                & 0.93 & 0.34 & 0.81 & 10.1 &   -  & 0.47 & 0.17\\

    \midrule
    \multirow{6}{*}{\textit{C-DiT}} 
        &  \clapa, \ctx   & 1.11 & 0.38 & 0.33 & 2.42 & 0.72 & 0.47 & 0.15\\
        &  \clapt, \ctx   & 1.13 & 0.39 & 0.27 & 3.15 & 0.73 & 0.43 & 0.15\\
        & \ctx            & 1.08 & 0.36 & 0.37 & 3.73 & 0.67 & 0.39 & 0.10\\
        & \clapa          & 1.27 & 0.43 & 0.25 & 2.01 &   -  & 0.44 & 0.11\\
        & \clapt          & 1.29 & 0.43 & 0.20 & 2.41 &   -  & 0.39 & 0.13\\
        & \uncond         & 1.23 & 0.41 & 0.33 & 2.83 &   -  & 0.37 & 0.07\\

    \bottomrule
    \multicolumn{9}{l}{$^{\mathrm{a}}$\(\times{10^{-3}}\); $^{\mathrm{b}}$\(\times{10^{-2}}\); $^c$ obtained from white noise}\\
    \multicolumn{9}{l}{$^d$ obtained using a random accompaniment from the dataset}
\end{tabular}
\end{footnotesize}
\end{center}
\vspace{-0.3cm}
 \caption{\small{Objective metrics obtained for each configuration under different conditional settings. We compare against high-performance bounds obtained from the \emph{real} validation set and low-performance bounds obtained from either \textit{noise} or randomly paired music contexts and accompaniments. Some cells are empty for APA in the case of context-free generation.}}
 \label{tab:results}

\end{table*}

Note that the values for the original Diff-A-Riff are different than in the original publication (cf. \cite{diffariff}). This is because here, we compare Diff-A-Riff generations with original audio data (e.g., to compute FAD, Coverage, Density, etc.), while in the original publication, all comparisons were done between generations and original data after a music2latent roundtrip (i.e., original data was encoded and decoded with music2latent). Also note that in the original Diff-A-Riff paper, the text conditioning was done using prompts generated by ChatGPT, while in this paper, we only use human-annotated lists of tags. Besides that, we use the same model architecture and weights as in the original publication.

Overall, the results improve when successively adding M2L2, DiT and $\textit{CLAP}_{\beta}$ (\clapa$^{*}$ is a special case, where the unconditioned Modality-Gap Bridge diffusion model is used to generate CLAP audio embeddings). The results for \textit{C-DiT} (i.e., the five-step latent consistency model) are slightly worse than with the conventional diffusion model but still substantially better than those of the original Diff-A-Riff model. We observe a slight decrease, though, in metrics such as APA and Clap Score ($CS_{TA}$). Consistency models are designed to generate samples in fewer steps by learning to approximate the denoising process more directly. However, this acceleration can come at the cost of reduced conditioning fidelity. One contributing factor is that consistency models cannot benefit as much from classifier-free guidance as traditional diffusion models. This limitation may lead to slightly lower alignment with the provided context and text prompts.

An interesting observation is that the best results in terms of KD$^a$, FAD, Density, and Coverage are achieved by $\textit{CLAP}_{\beta}$ using text conditioning (\clapt). We hypothesize that this improvement comes from generating audio CLAP embeddings from text CLAP embeddings via our diffusion model $\textit{CLAP}_{\beta}$, which allows us to sample from a more diverse audio embedding distribution space and effectively transforms (unseen) text embeddings into audio embeddings that are within the conditional distribution the generative model was trained on. Consequently, this leads to improved performance on metrics that assess the quality and coverage of the generated samples.

The APA metric \cite{apa} reflects how well the generated audio aligns with the given audio context. The \textit{+DiT} models achieve the best results in APA, demonstrating that these models are highly responsive to the provided audio context in generating accompaniments. Also, the new variants overall have a much better APA than the original Diff-A-Riff when only context conditioning is provided (\ctx). This shows that the efficacy of the architecture improved considerably and that additional cues are not needed to produce valid accompaniments. This independence of conditionings is also striking when considering the improvements for fully unconditional generation (\textit{uncond}).

When considering text-derived CLAP embeddings ($\textit{CLAP}_{T}$), the Clap Score ($CS_{TA}$) measures the alignment between the text prompts and the generated audio. The original Diff-A-Riff attains a $CS_{TA}$ of 0.23 with both $\textit{CLAP}_{T}$ and Context, and 0.24 with $\textit{CLAP}_{T}$ only. Our enhanced models show a slight decrease in CS under the same settings. A possible reason for that is the fact that in the original Diff-A-Riff training, the CLAP audio embedding was computed from the exact target segment, while here, we compute the embedding from a random segment of the target stem. This means that the new model variants generalize better to the instrument type and are less sensitive to the exact position of the CLAP embedding. Due to the gap between text and audio embeddings in CLAP, the generated audio is, therefore, less likely to project back exactly to the text embedding.

\vspace{-0.2cm}
\section{Conclusion}
\label{sec:conclusion}

This work presents a series of substantial enhancements to Diff-A-Riff, a state-of-the-art latent diffusion model for musical accompaniment co-creation. By integrating a higher-fidelity stereo autoencoder, a transformer-based diffusion architecture, an advanced diffusion framework, and consistency training, we achieve significant improvements in audio quality, diversity, and inference speed. Additionally, our novel method for bridging the modality gap in CLAP embeddings enhances the model's responsiveness to text prompts, paving the way for more intuitive and precise text-driven audio generation. Through rigorous objective evaluation and ablation studies, we demonstrate the effectiveness of our proposed enhancements, highlighting the potential of our model for practical applications in music production. This work represents a further step towards developing AI-assisted tools that empower musicians with enhanced creative control and facilitate seamless integration of machine-generated content into their artistic workflows.




\printbibliography








\end{document}